\documentclass{PoS}
\usepackage{macros,macros_static}
\usepackage{amstext,amsfonts,amsbsy,amsmath,amssymb,amscd,epsf,mathrsfs,dsfont}
\usepackage{rotating}
\usepackage{multicol}
\usepackage{pstricks}
\usepackage{pst-node}
\usepackage{latexsym}
\usepackage{epsfig}
\usepackage{ulem} 
\def\Oa{O($a$) }
\def\mpi2{m_\pi^2}
\def\cQ{{\cal Q}}
\def\cZ{{\cal Z}}
\def\cA{{\cal A}}

\newcommand{\msbar}{{\overline{\rm MS}}}

\newcommand{\bflr}{\begin{flushright}}
\newcommand{\eflr}{\end{flushright}}
\newcommand{\bfll}{\begin{flushleft}}
\newcommand{\efll}{\end{flushleft}}
\newcommand{\ben}{\begin{enumerate}}
\newcommand{\een}{\end{enumerate}}
\newcommand{\beq}{\begin{equation}}
\newcommand{\eeq}{\end{equation}}
\newcommand{\beqn}{\begin{eqnarray}}
\newcommand{\eeqn}{\end{eqnarray}}

\title{Non-perturbative renormalization and running of $\Delta F=2$ four-fermion operators in the SF scheme}

\ShortTitle{Non-perturbative renormalization of $\Delta F=2$ four-fermion operators}

\author{\speaker{Mauro Papinutto}\\
        Dipartimento di Fisica, "Sapienza" Universit\`a di Roma, and INFN, Sezione di Roma,\\ 
        Piazzale Aldo Moro 2, I-00185 Roma, ITALY.\\
        E-mail: \email{mauro.papinutto@roma1.infn.it}}

\author{Carlos Pena, David Preti\\
             Instituto de F\'\i sica Te\'orica UAM/CSIC and Departamento de F\'\i sica Te\'orica, Universidad
Aut\'onoma de Madrid, Cantoblanco E-28049 Madrid, Spain\\
         E-mail: \email{carlos.pena@uam.es, david.preti@csic.es}}

\abstract{We present preliminary results of a non-perturbative study of the scale-dependent renormalization constants of a complete basis of $\Delta F=2$ parity-odd four-fermion operators that enter the computation of hadronic B-parameters within the Standard Model (SM) and beyond. We consider non-perturbatively \Oa improved Wilson fermions and our gauge configurations contain two flavors of massless sea quarks. The mixing pattern of these operators is the same as for a regularization that preserves chiral symmetry, in particular there is a "physical" mixing between some of the operators. The renormalization group running matrix is computed in the continuum limit for a family of Schr\"odinger Functional (SF) schemes through finite volume recursive techniques. We compute non-perturbatively the relation between the renormalization group invariant operators and their counterparts renormalized in the SF at a low energy scale, together with the non-perturbative 
matching matrix between the lattice regularized theory and the various SF schemes.}

\FullConference{The 32nd International Symposium on Lattice Field Theory,\\
		23-28 June, 2014\\
		Columbia University New York, NY}

\begin{document}

\section{Introduction}
\vspace{-0.2cm}

Flavor physics processes play a major role in the indirect search for New Physics (NP), because they are sensitive to the exchange of virtual NP particles through loop effects. These processes vanish at tree level in the SM and, despite the fact they are loop mediated and in some cases also CKM or helicity suppressed, may be theoretically very clean. Among them $\Delta F=2$ transitions have always provided some of the most stringent constraints on NP. 


The most general $\Delta F=2$ weak effective Hamiltonian beyond the SM can be written in terms of the following parity even (PE) and parity odd (PO) four fermion operators :
\bea
Q_1^\pm = \bar f\gamma_\mu q {\bar f}  \gamma_\mu q' +  \bar f\gamma_\mu\gamma_5 q {\bar f}  \gamma_\mu\gamma_5 q'  \pm (q\leftrightarrow q') &\qquad& {\cal Q}_1^\pm = \bar f\gamma_\mu q {\bar f}  \gamma_\mu\gamma_5 q' +  \bar f\gamma_\mu\gamma_5 q {\bar f}  \gamma_\mu q'  \pm (q\leftrightarrow q')\nonumber\\
Q_2^\pm = \bar f\gamma_\mu q {\bar f}  \gamma_\mu q' -  \bar f\gamma_\mu\gamma_5 q {\bar f}  \gamma_\mu\gamma_5 q'  \pm (q\leftrightarrow q') && {\cal Q}_2^\pm = \bar f\gamma_\mu q {\bar f}  \gamma_\mu\gamma_5 q' -  \bar f\gamma_\mu\gamma_5 q {\bar f}  \gamma_\mu q'  \pm (q\leftrightarrow q')\nonumber\\
Q_3^\pm = \bar f q {\bar f} q' -  \bar f \gamma_5 q {\bar f}  \gamma_5  q'  \pm (q\leftrightarrow q')\qquad && \qquad {\cal Q}_3^\pm = \bar f \gamma_5 q {\bar f} q' -  \bar f q {\bar f} \gamma_5 q'  \pm (q\leftrightarrow q')\nonumber\\
Q_4^\pm = \bar f q {\bar f} q' + \bar f \gamma_5 q {\bar f}  \gamma_5  q'  \pm (q\leftrightarrow q')\qquad && \qquad {\cal Q}_4^\pm = \bar f \gamma_5 q {\bar f} q' +  \bar f q {\bar f} \gamma_5 q'  \pm (q\leftrightarrow q')\nonumber\\
Q_5^\pm = 2 \bar f \sigma_{\mu\nu} q {\bar f}\sigma_{\mu\nu} q'  \pm (q\leftrightarrow q')\qquad \quad\ && \qquad  {\cal Q}_5^\pm =2 \bar f \sigma_{\mu\nu}\gamma_5 q {\bar f}\sigma_{\mu\nu}\gamma_5 q'  \pm (q\leftrightarrow q')\nonumber
\eea
where the flavours $q$ and $q'$ can be thought of as two copy of the same flavour and will be set to be identical in the end. This allows to define "$^+$" and "$^-$" operators which are even or odd under the switching symmetry $q\leftrightarrow q'$ and thus do not mix with each other. In the following we will drop the superscript "$^\pm$" and our discussion will apply to both "$^+$" and "$^-$" sectors separately. Moreover, parity symmetry prevents PE operators from mixing with PO ones. Finally we note that the operators $Q_2, Q_3, Q_4, Q_5, \cQ_2, \cQ_3, \cQ_4, \cQ_5$ appear only in extensions beyond the SM. 

In a regularisation that preserves chiral symmetry, $Q_2$ mixes under renormalization with $Q_3$, and similarly $Q_4$ with $Q_5$. The same mixing pattern is valid for $\cQ_2$ and $\cQ_3$ and for $\cQ_4$ and $\cQ_5$, which share the same chiral properties of the corresponding PE operators.
The corresponding $2\times2$ renormalization matrices (we will call each of them $Z$ and $\cZ$) have large LO anomalous dimensions, and the same is true at NLO (even though this is a scheme-dependent statement). This poses some issues about the use of perturbation theory (PT) to compute the renormalization factors and/or the Renormalization Group (RG) running down to renormalization scales at which the operators are usually renormalized on the lattice. 

At present, few computations of these matrix elements in the $\Delta S=2$ case exist with dynamical fermions~\cite{Boyle:2012qb,Bertone:2012cu,Bae:2013tca}. The first two works~\cite{Boyle:2012qb,Bertone:2012cu} use non-perturbative renormalization in the RI-MOM scheme at scales of $2-3\ \GeV$, and perturbative 
RG-running at NLO, while Ref.~\cite{Bae:2013tca} uses perturbative renormalization and running. While the results from~\cite{Boyle:2012qb,Bertone:2012cu} are roughly consistent, there are substantial discrepancies with those of~\cite{Bae:2013tca}, and this could point to a systematic error in the use of perturbative renormalization. Furthermore, in view of the large anomalous dimensions mentioned above, the use of perturbative running starting from scales of $2-3\ \GeV$ could also represent a substantial source of systematic error for all of the previous three computations. The aim of the present study is to investigate non-perturbatively the RG running from a hadronic scale $\mu_{\rm had}\sim \Lambda_{QCD}$ up to a scale $\mu_{\rm pt}$ of the order of the $W$ boson mass $M_W$, where matching with PT at NLO is under much better control.

\begin{figure}[!ht]
\label{fig:correlators}
\vspace{0.25cm}
\begin{center}
\epsfig{figure=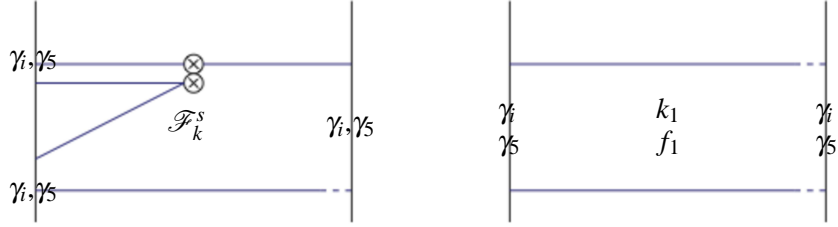,angle=0,width=0.70\linewidth}
\put(-250,35){${\cal F}_k^{s}$}
\put(-310,60){$\gamma_i$,$\gamma_5$}
\put(-310,10){$\gamma_i$,$\gamma_5$}
\put(-190,35){$\gamma_i$,$\gamma_5$}
\put(-65,28){$f_1$}
\put(-5,28){$\gamma_5$}
\put(-125,28){$\gamma_5$}
\put(-65,40){$k_1$}
\put(-5,40){$\gamma_i$}
\put(-125,40){$\gamma_i$}
\end{center}
\vspace{-0.5cm}
\caption{Left: 4-point correlators ${\cal F}_k^{s}$ of the operator $\cQ_k$ with the source $s$. Right: boundary-to-boundary correlators $k_1$ ($\gamma_i$ on the boundaries) and $f_1$ ($\gamma_5$ on the boundaries).}
\vspace{-0.1cm}
\end{figure}

Due to the explicit breaking of chiral symmetry with Wilson fermions, the renormalization pattern of composite operators can be considerably more complex than in a chirality-preserving regularisation, because of the mixing with operators of different na\"\i ve chirality. While the 5 PE operators $Q_i$ all mix with each other, it has been shown~\cite{Donini:1999sf} that in the PO sector the mixing pattern is the same as in a chirality-preserving regularisation, i.e. the one described above.

As a consequence one can think of two possible strategies to avoid the spurious mixing in the PE sector when using non-perturatively \Oa improved Wilson sea-quarks:
\bi
\item use twisted mass QCD at maximal twist for $f$, $q$ and $q'$ as explained in~\cite{Frezzotti:2004wz}. This setup is automatically \Oa improved but not unitary;
\item use Ward identities which relate the correlators of PE operators to those of PO ones as explained in~\cite{Becirevic:2000cy}. This setup is unitary but not automatically \Oa improved.
\ei
In both strategies one has to compute the renormalized matrix elements of PO operators which present only the "physical" scale dependent mixing. 

In the present work we focus on the two matrices $k=2,3$ and $k=4,5$, the renormalization of the operator $k=1$ having been already studied in~\cite{Palombi:2005zd}. This is the first time that the RG-running in the presence of mixing has been computed non-perturbatively over a wide range of scales $\mu\in[\Lambda_{QCD}, M_{W}]$. In this exploratory study we have used non-perturbatively \Oa improved Wilson fermions with 2 massless sea flavors.

\section{Non-perturbative renormalization in the SF scheme}
By using the SF on a volume $L^4$ we compute the 4-point correlators ${\cal F}_k^{s}$ of the operator $\cQ_k$ with the source $s$ made by one of the five possible combinations of three $\gamma_i$ and $\gamma_5$ bilinears on the boundaries. We also compute the correlators of two boundary bilinears with a $\gamma_5$ structure ($f_1$) or with a $\gamma_i$ structure ($k_1$), see~\cite{Palombi:2005zd} for details. These correlators are schematically represented in Fig.~\ref{fig:correlators}

From the correlators we build the ratios 
\bea
\cA_{k;\alpha}^{s}(L/2) = \frac{ {\cal F}_k^{s}(L/2)}{f_1^{3/2-\alpha}k_1^{\alpha}}\ ,
\eea
where $\alpha\in\{0,1,3/2\}$ and $s\in\{1,\ldots,5\}$. We impose renormalization conditions in the chiral limit (i.e. at bare mass $m_0=m_{\rm cr}$) on each of the $2\times2$ matrices by choosing two combinations of sources $s_1$,$s_2$ for each value of $\alpha$ (to simplify the notation we avoid labelling $\cZ$ with the indices $s_1$,$s_2,\alpha$):
\beqn
 \left(\begin{array}{cc}
	  \cZ_{22} & \cZ_{23}   \\
	  \cZ_{32} & \cZ_{33}  
	\end{array}\right)
\left(\begin{array}{cc}
	  \cA^{s_1}_{2;\alpha} & \cA^{s_2}_{2;\alpha}   \\
	  \cA^{s_1}_{3;\alpha} & \cA^{s_2}_{3;\alpha}  
	\end{array}\right)
	=
\left(\begin{array}{cc}
	  \cA^{s_1}_{2;\alpha} & \cA^{s_2}_{2;\alpha}   \\
	  \cA^{s_1}_{3;\alpha} & \cA^{s_2}_{3;\alpha}  
	\end{array}\right)_{\black {g^2_0=0}}
\eeqn
and similarly for the $k=4,5$ matrix. In order for the renormalization condition defined by $(s_1,s_2,\alpha)$ to make sense, one needs to check that
\bea
\det \left(\begin{array}{cc}
	  \cA^{s_1}_{2;\alpha} & \cA^{s_2}_{2;\alpha}   \\
	  \cA^{s_1}_{3;\alpha} & \cA^{s_2}_{3;\alpha}  
	\end{array}\right)_{\black {g^2_0=0}} \neq 0
\eea
and similarly for the $k=4,5$ matrix. It turns out that the only non-redundant renormalization conditions correspond to the 6 choices $(s_1,s_2)\in\{(1,2),(1,4),(1,5),(2,3),(3,4),(3,5)\}$\footnote{S.~Sint, unpublished notes, 2001}. Considering the fact that we allow for three values of $\alpha$, we have in all 18 non-redundant conditions. Each of these conditions fixes the two renormalization matrices $\cZ(g_0,a\mu)$ at the scale $\mu=1/L$.

From the renormalization constants we build the Step Scaling Functions (SSF) for the two matrices $\cZ$ (where we recall also the definition for the SSF of the coupling and where $U(\mu_1,\mu_2)$ is the RG-evolution between the scales $\mu_1$ and $\mu_2$):
\beqn
\Sigma_{\cQ}(u,a/L)&\equiv& \cZ(g_0,2L/a)\left[\cZ(g_0,L/a)\right]^{-1}
  \Big|_{u\equiv\gbar_{\rm SF}^2(L)}^{m=m_{\rm cr}}\ ,\nonumber\\
 \sigma_{\cQ}(u) &=& U(\mu/2,\mu)\Big|_{\mu=1/L}=\lim_{a \to 0} \Sigma_{\cQ}(u,a/L)\ ,\\
 &\sigma(u)&\equiv\gbar_{\rm SF}^2(2L)\ ,\qquad u\equiv\gbar_{\rm SF}^2(L)\ .\nonumber
\eeqn
The SSF $\sigma(u)$ has been computed in previous works by the ALPHA Coll. We have computed here the two matrices $\sigma_{\cQ}(u)$ ($k=2,3$ and $k=4,5$) for the 18 schemes and for 6 values of the coupling in the range $u=0.9793$ to $u=3.3340$.

Our operators are not \Oa improved, so the continuum limit extrapolation is linear in $a/L$ and has been performed using lattices with $L/a=\{6,8,12\}$ and $2L/a=\{12,16,24\}$
by tuning the $\beta$ values at each $L$ to obtain the chosen value of $u$.
 
Having the continuum limit of $\sigma_{\cQ}(u)$ for $u\in[0.9793,3.3340]$, we can perform a 
fit according to a power series expansion $\sigma_{\cQ}(u)=\mathbf{1} + s_1 u +  s_2 u^2 +  s_3 u^3 + \ldots$, where the $s_i$ are $2\times2$ matrices. In perturbation theory they can be related to the coefficients of the anomalous dimension matrix (ADM) and beta function
\beqn
\gamma(g)=-g^2\sum_{n=0}^{\infty}\gamma_n g^{2n}\ , &&\qquad\qquad \beta(g)=-g^3\sum_{n=0}^{\infty}\beta_n g^{2n}\ ,\nonumber\\
s_1=\gamma_0 \ln 2\ ,\qquad\qquad && s_2=\gamma_1 \ln 2 +\beta_0\gamma_0(\ln 2)^2 +\frac{1}{2}\gamma_0^2 (\ln 2)^2\ , 
\label{eq:s1s2}
\eeqn
where $\gamma_1$ is the NLO ADM in the SF scheme (we denote it by $\gamma_1^{\rm SF}$). The latter can be obtained from the value $\gamma^{\rm ref}_1$ already known in a reference scheme through the following two-loop matching relations:
\vspace*{-0.5cm}
\beqn
\gamma^{\rm SF}_1=\gamma_1^{\rm ref}+[\chi_{\rm SF,ref}^{(1)},\gamma_0]&+&2\beta_0\chi_{\rm SF,ref}^{(1)}+\beta_0^\lambda\lambda\frac{\partial}{\partial \lambda}\chi_{\rm SF,ref}^{(1)}-\gamma^{(0)}\chi_g^{(1)}\ ,\nonumber\\
\gbar^2_{\rm SF}=\chi_g(g_{\rm ref})g^2_{\rm ref}\ ,&&\qquad(\cQ)^{\rm SF}_{\rm R} 
= \chi_{\rm SF,ref}(g_{\rm ref}) (\cQ)^{\rm ref}_{\rm R}\ ,\\
\chi(g)&=&1+\sum_{k=1}^{\infty}g^{2k}\chi^{(k)}\ ,\nonumber
\eeqn
where $\lambda$ is the gauge fixing parameter and $\beta_\lambda$ is the beta function for the renormalized gauge fixing parameter $\lambda(\mu)$ (This is needed e.g. if we use as reference scheme the RI-MOM which depends on the gauge chosen. If we use $\MSbar$ there is no dependence upon the gauge.)

\begin{figure}[!ht]
\vspace*{-0.7cm}
\begin{center}
      \hskip -0.5cm\includegraphics[width=5.1 true cm,angle=0]{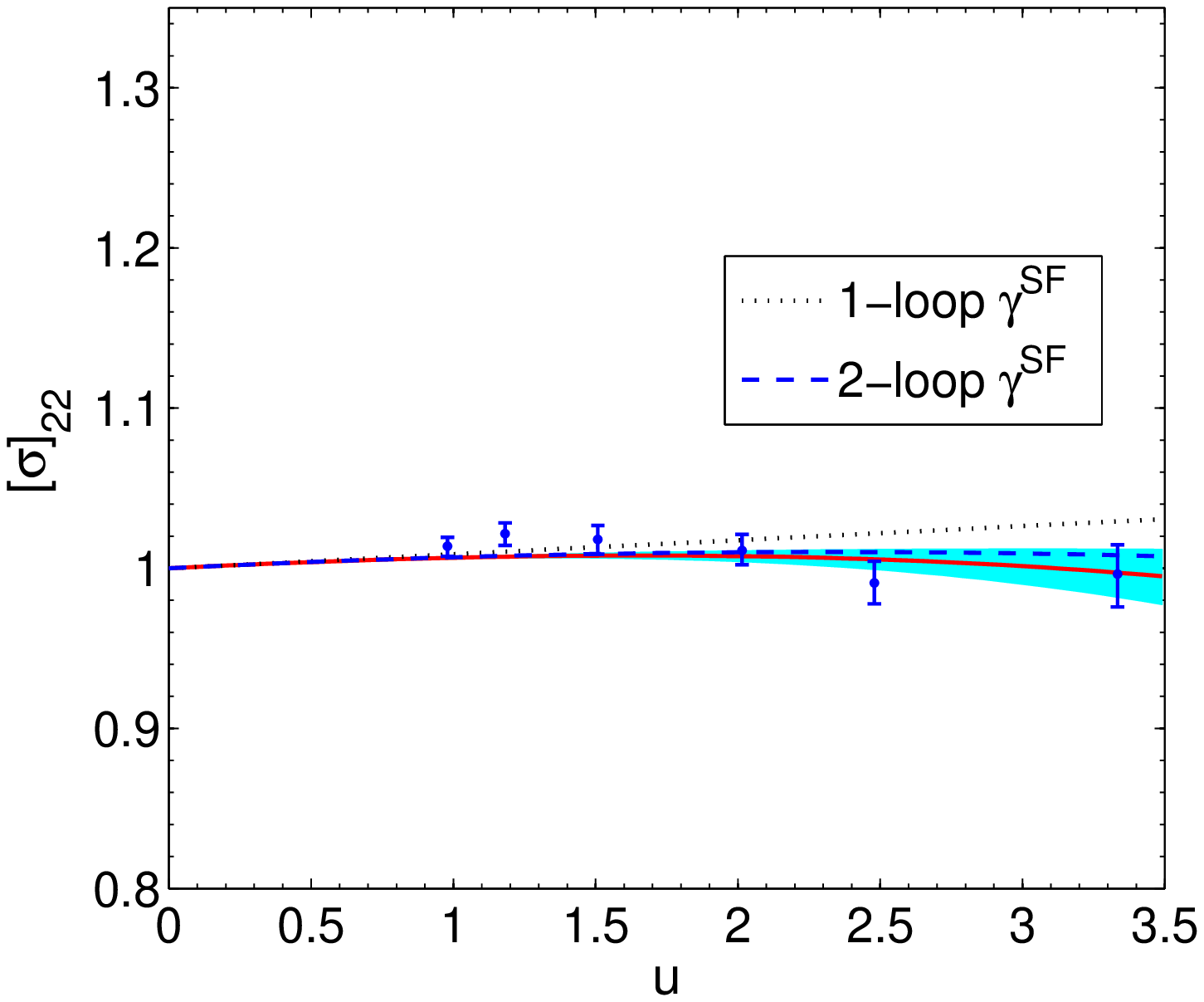}
      \hskip 0.0cm \includegraphics[width=5.1 true cm,angle=0]{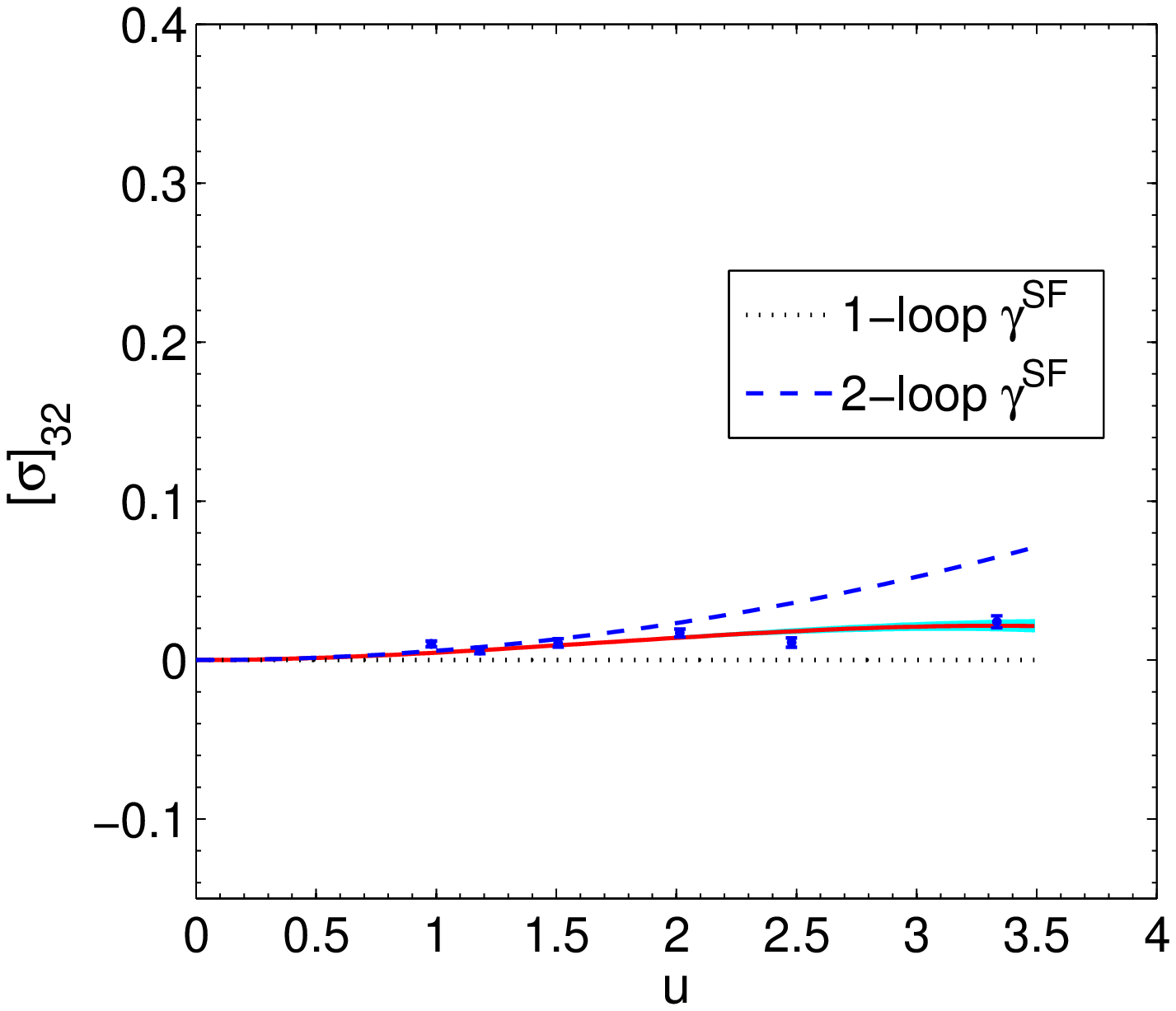}
      \hskip 0.0cm\includegraphics[width=5.1 true cm,angle=0]{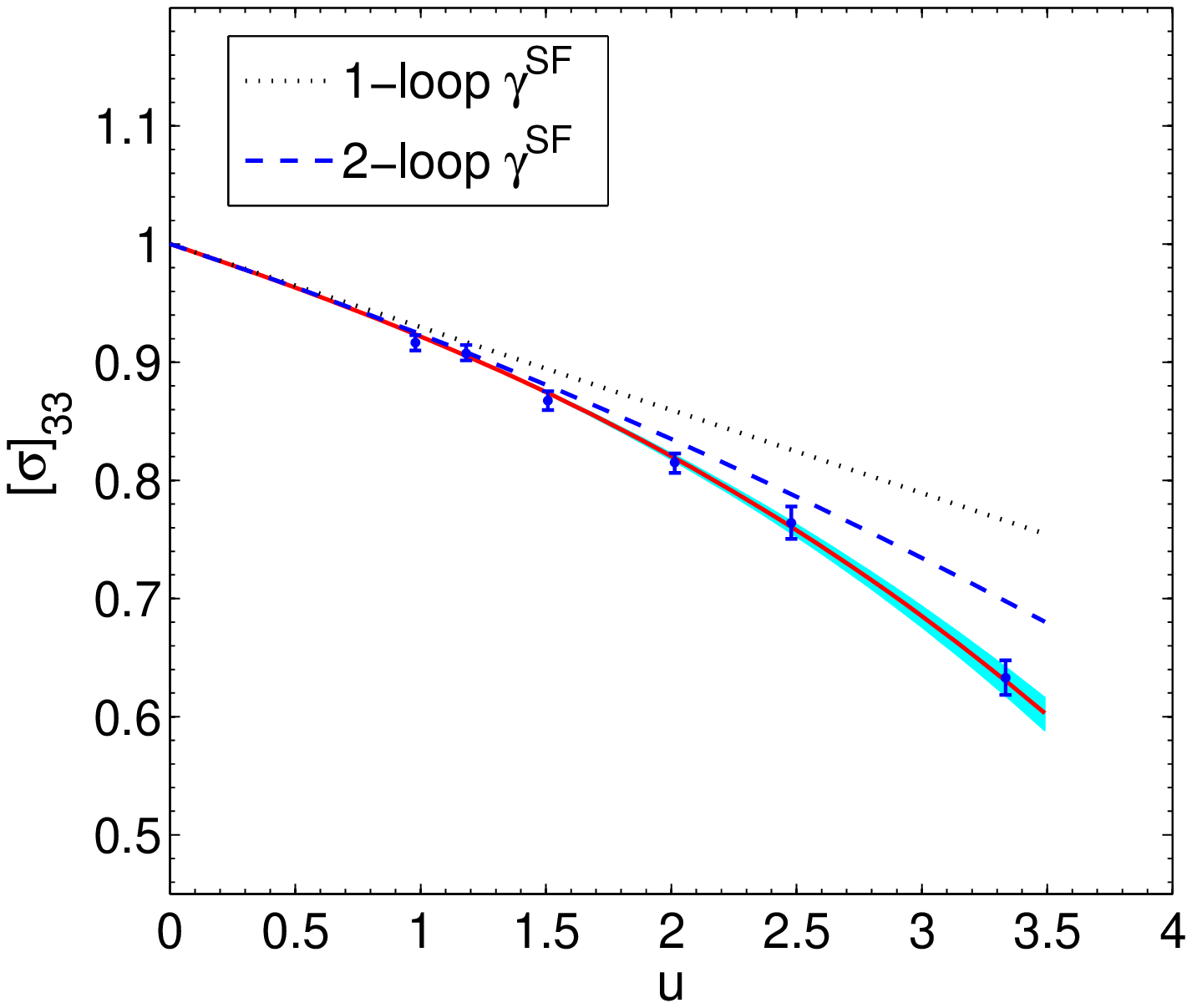}
     \vskip 0.0cm
     \hskip -0.5cm\includegraphics[width=5.1 true cm,angle=0]{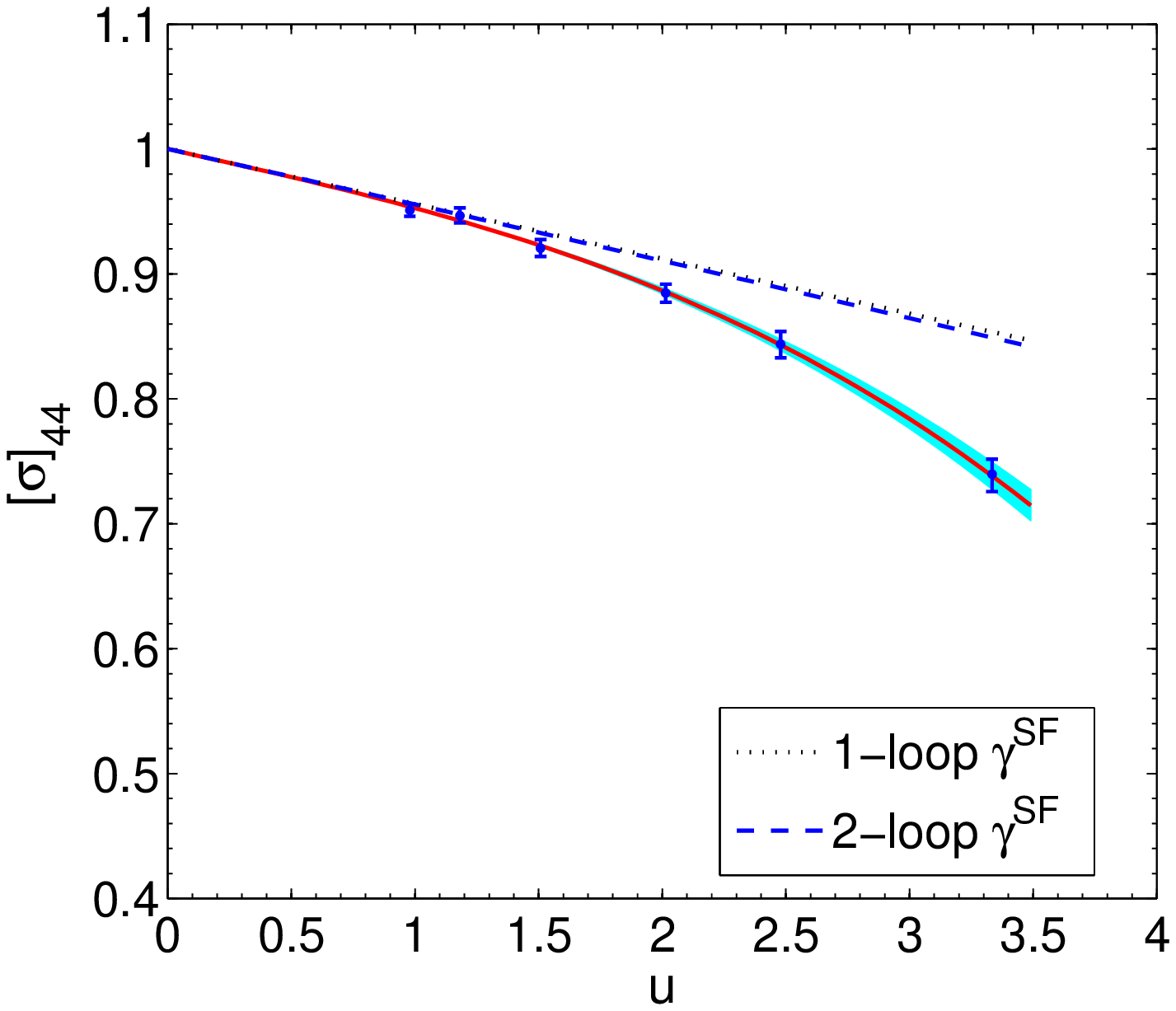}
      \hskip 0.0cm\includegraphics[width=5.1 true cm,angle=0]{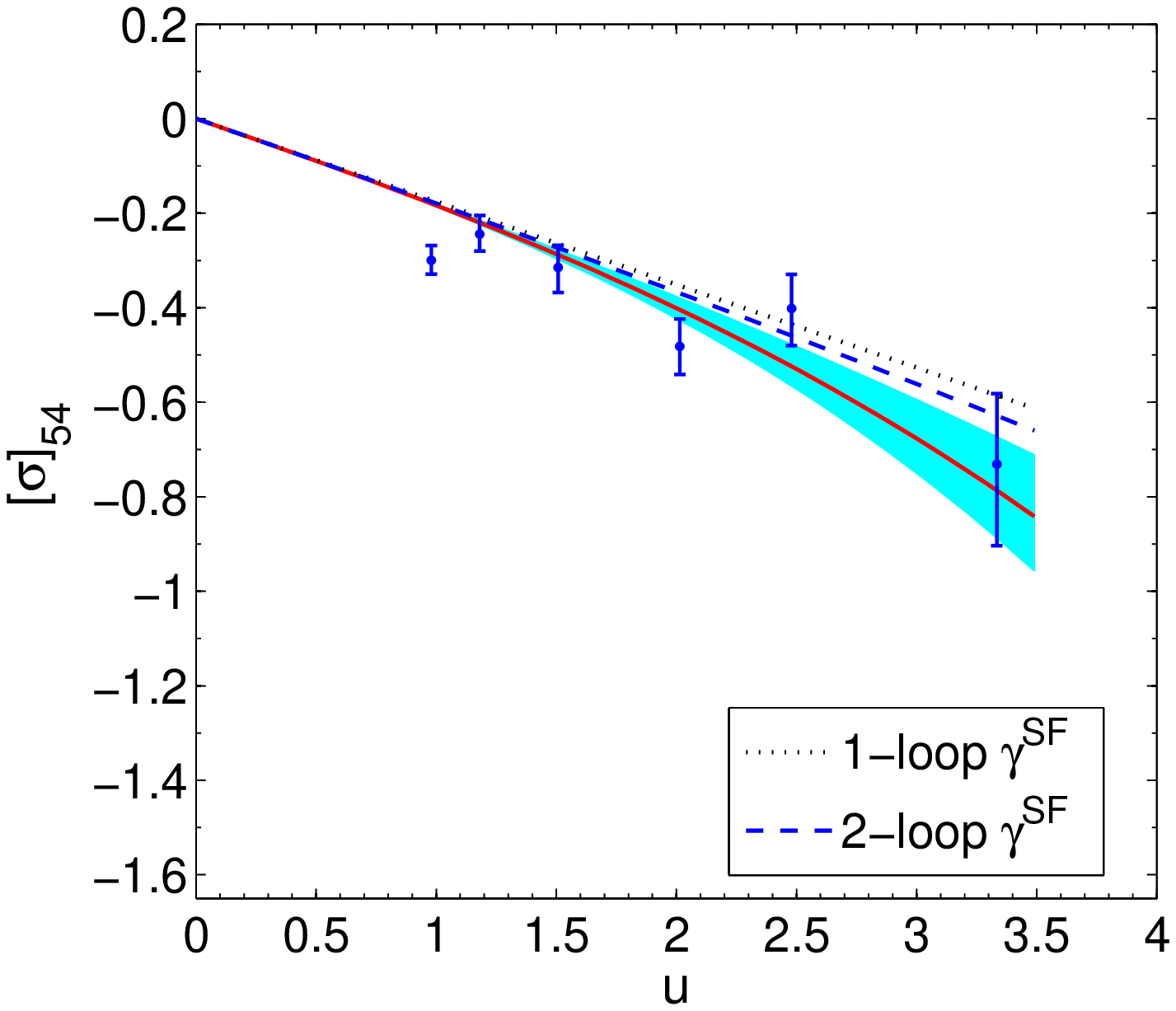}
      \hskip 0.0cm \includegraphics[width=5.1 true cm,angle=0]{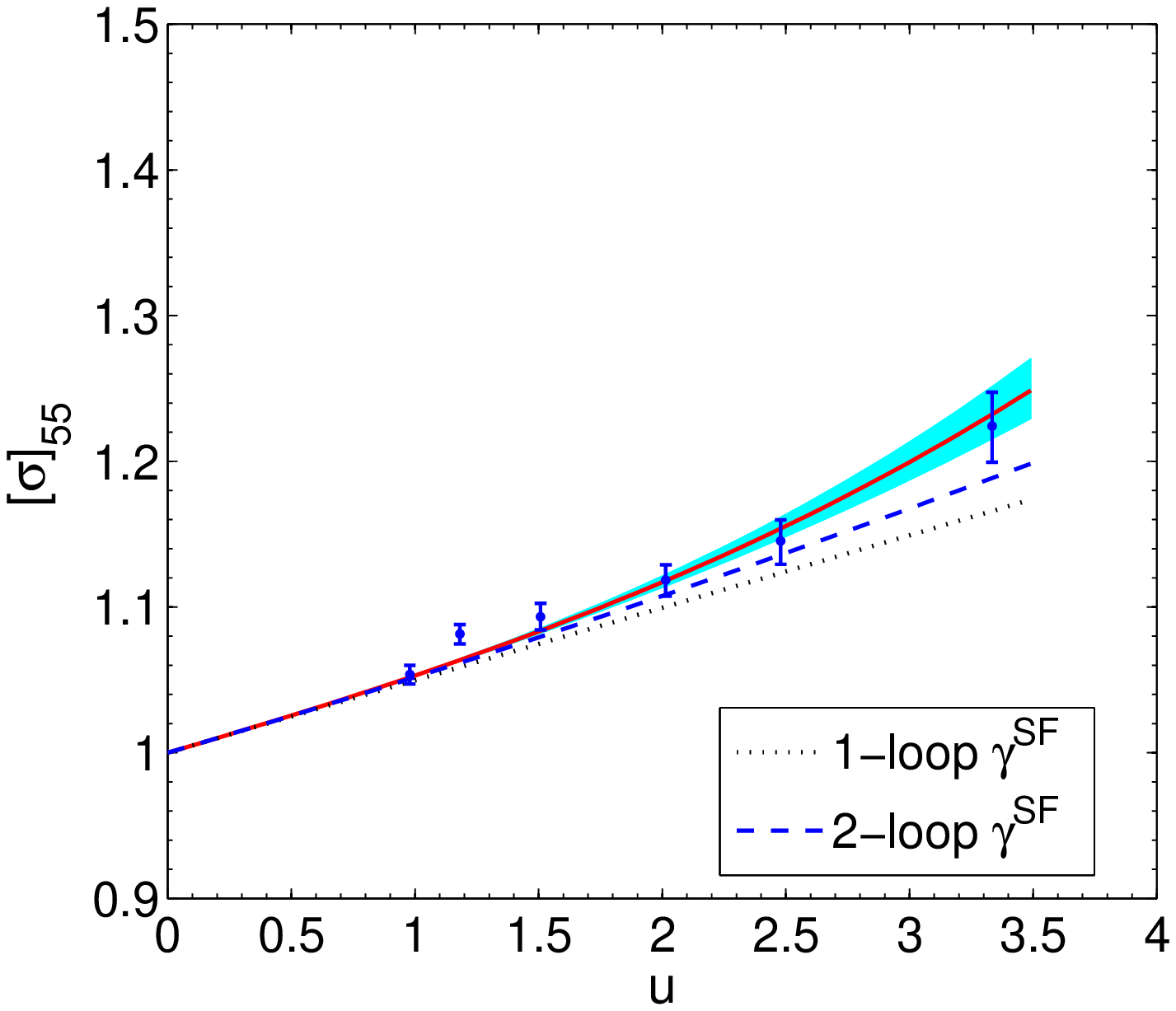}
      \vskip -0.3cm
    \caption{First row: elements 22, 32, 33 of $\sigma_{\cQ}(u)$. Second row: 
    elements 44, 54, 55 of $\sigma_{\cQ}(u)$.}
    \label{fig:ssf}
  \end{center}
\vskip -0.6cm
\end{figure}

The matching coefficient $\chi_{\rm SF,ref}^{(1)}=\chi^{(1)}_{\rm SF,lat}-\chi^{(1)}_{\rm ref,lat}$ involves the respective one-loop matching matrices $\chi^{(1)}_{\rm SF,lat}$ and $\chi^{(1)}_{\rm ref,lat}$ between the bare lattice operator and either the SF or the reference scheme. In the present work we have computed the matching matrix $\chi^{(1)}_{\rm SF,lat}$ in perturbation theory at one-loop. $\chi^{(1)}_{\rm ref,lat}$ can be extracted from the literature (Ref.~\cite{Constantinou:2010zs} for the RI-MOM scheme while Refs.~\cite{Gupta:1996yt,Kim:2014,Capitani:2001} for the $\MSbar$ scheme\footnote{We are grateful to S.~Sharpe for having converted for us the $\MSbar$ scheme used in~\cite{Gupta:1996yt} to the one defined in~\cite{Buras:2000if}.}).

$\gamma^{\rm ref}_1$ can be found in~\cite{Buras:2000if} both in the RI-MOM and $\MSbar$ case,
while $\chi^{(1)}_g$ is given in~\cite{Sint:1995ch}. Having all these ingredients we have computed $\gamma_1^{\rm SF}$, for both $2\times2$ matrices. A strong check is represented by the fact 
that the results obtained by using as reference scheme RI-MOM agree with those obtained by using the $\MSbar$ scheme.

From $\gamma^{\rm SF}_1$ we can easily compute $s_1$ and $s_2$ from Eq.~\ref{eq:s1s2} and then perform a fit of the two matrices $\sigma_{\cQ}(u)$ by keeping $s_3$ as a matrix of free parameters. 
As an example, results for some elements of $\sigma_{\cQ}(u)$ in the $\alpha=3/2$, $(s_1,s_2)=(1,5)$ scheme are presented in Fig.~\ref{fig:ssf}. In general, several of these elements differ substantially at the largest couplings from the LO and NLO PT results, independently of the scheme chosen.

\vspace*{-0.2cm}
\section{Non-perturbative renormalization group running}
\vspace*{-0.2cm}

Once the $\sigma_{\cQ}$ has been fitted on the whole range of couplings,
the non-perturbative running can be obtained from the scale $\mu_{\rm had}=1/L_{\rm max}$ to the scale 
$\mu_{\rm pt}=2^{n}\mu_{\rm had}$ where $n$ is the number of steps performed and where $L_{\rm max}$ is such that $\sigma^{-1}(\gbar^2(L_{\rm max}))$ belongs to the upper end of the range of couplings simulated:
\vspace*{-0.1cm}
\beqn
U(\mu_{\rm pt},\mu_{\rm had})=\big[\sigma_{\cQ}(u_1)\cdots\sigma_{\cQ}(u_{n})\big]^{-1},\qquad u_i=\gbar^2(2^{i}\mu_{\rm had})\ .
\eeqn
In the present case, with 7 steps we have $\mu_{\rm had}\approx 0.49$ GeV while 
$\mu_{\rm pt}\approx 63$ GeV, where one expects to safely match with the perturbative 
RG-evolution at NLO.

\begin{figure}[!ht]
\vspace*{-0.7cm}
\begin{center}
      \hskip -0.4cm\includegraphics[width=5.1 true cm,angle=0]{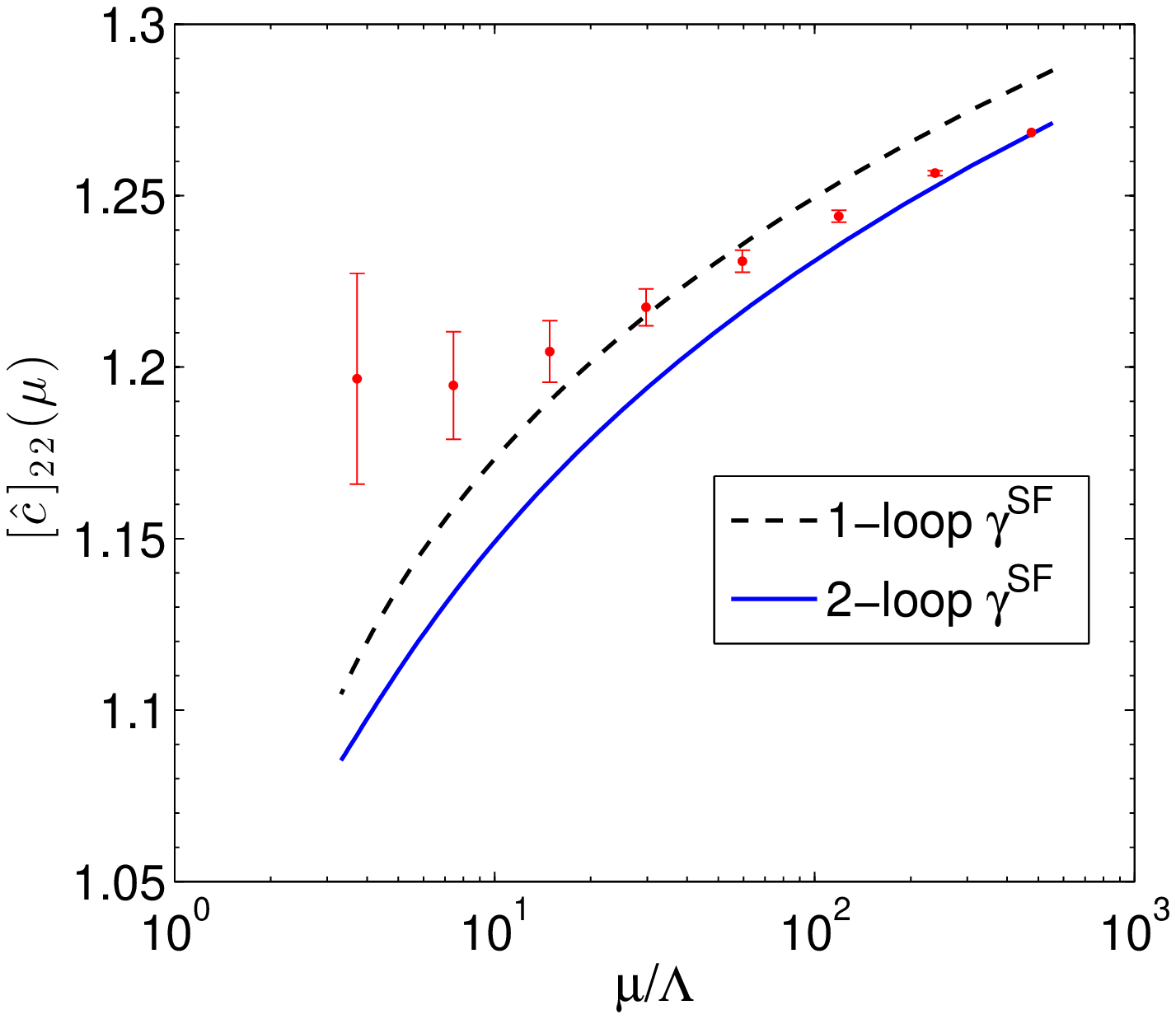}
      \hskip 0.0cm \includegraphics[width=5.1 true cm,angle=0]{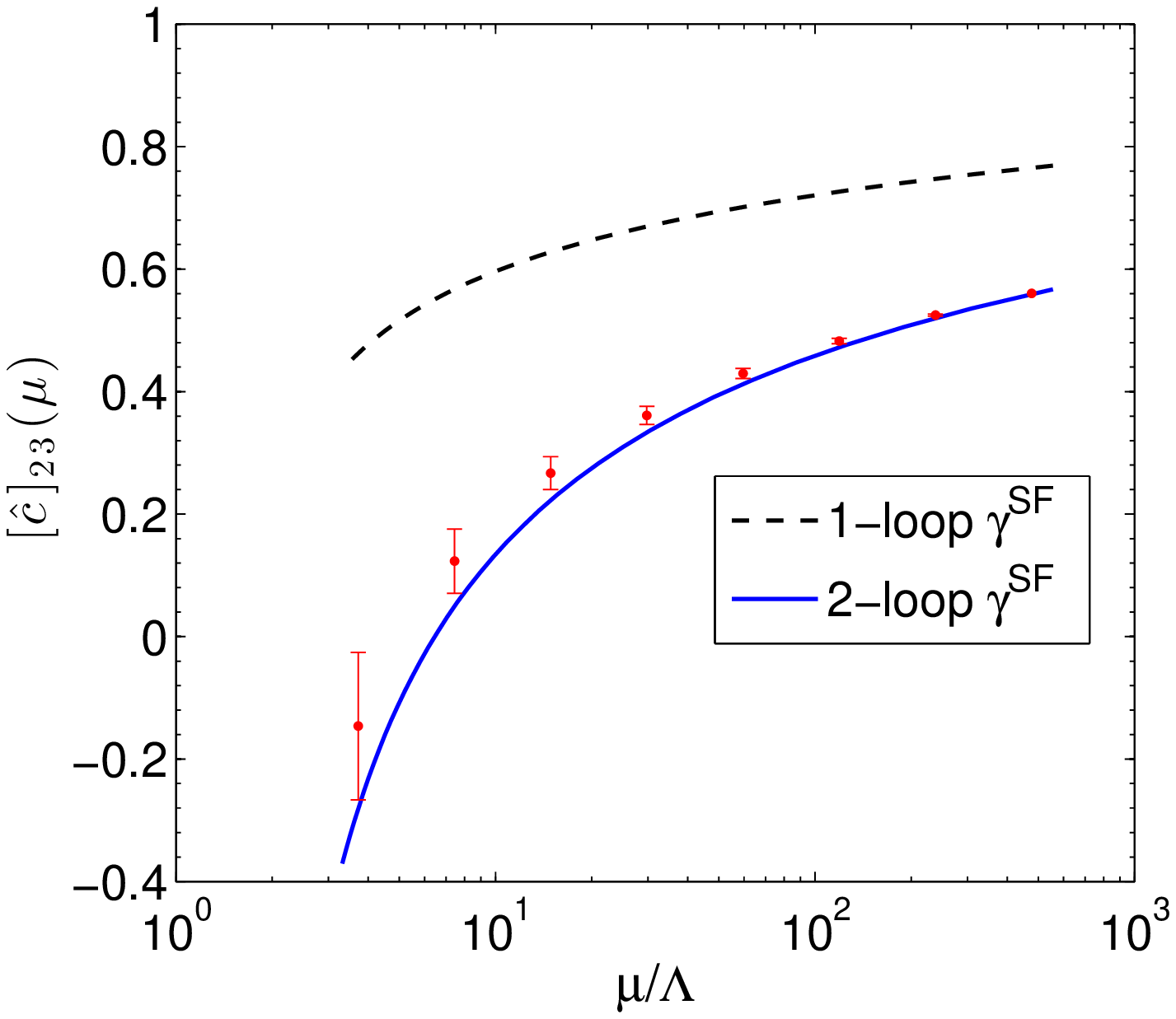}
      \hskip 0.0cm\includegraphics[width=5.1 true cm,angle=0]{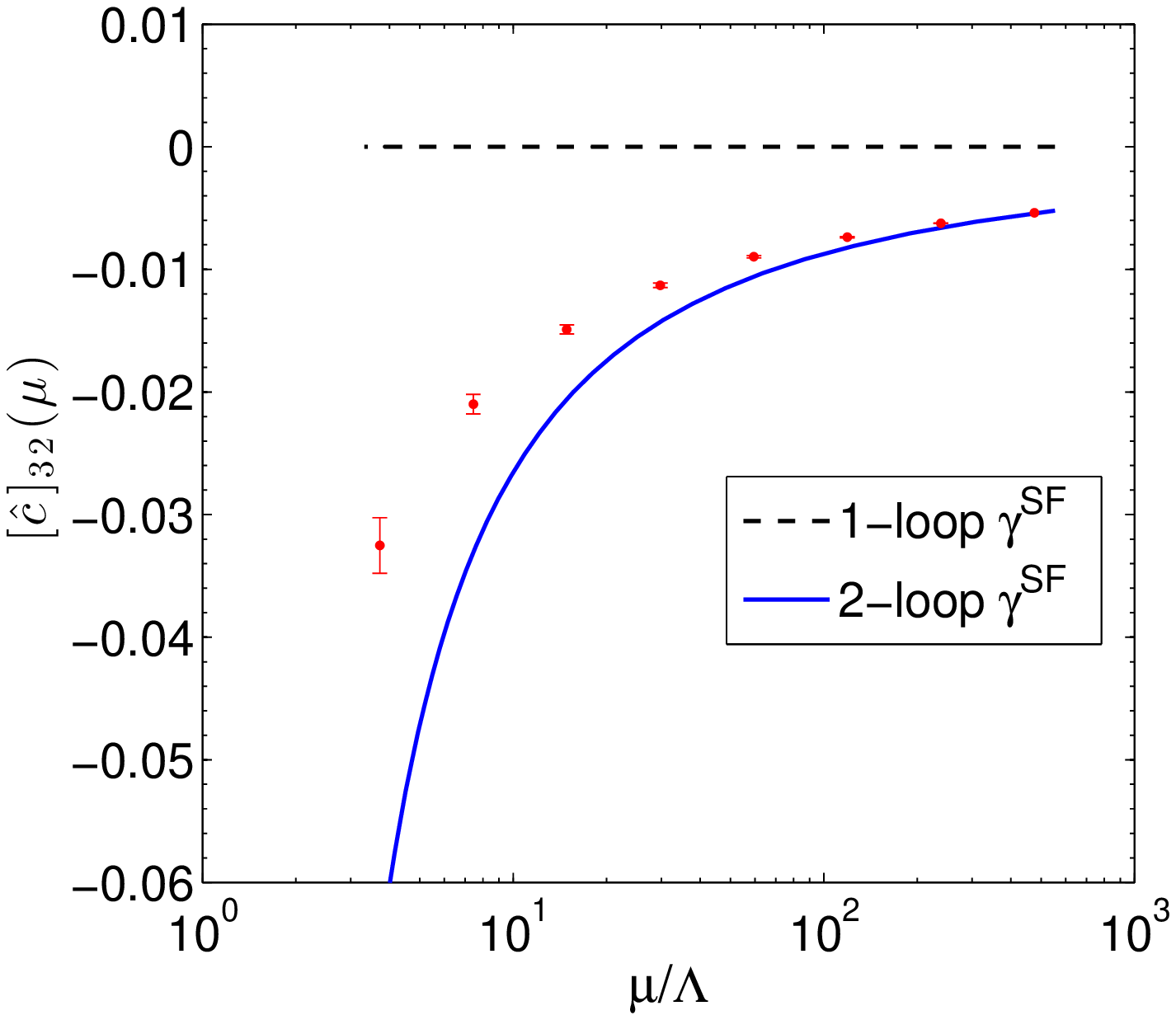}
     \vskip 0.0cm
     \hskip -0.4cm\includegraphics[width=5.1 true cm,angle=0]{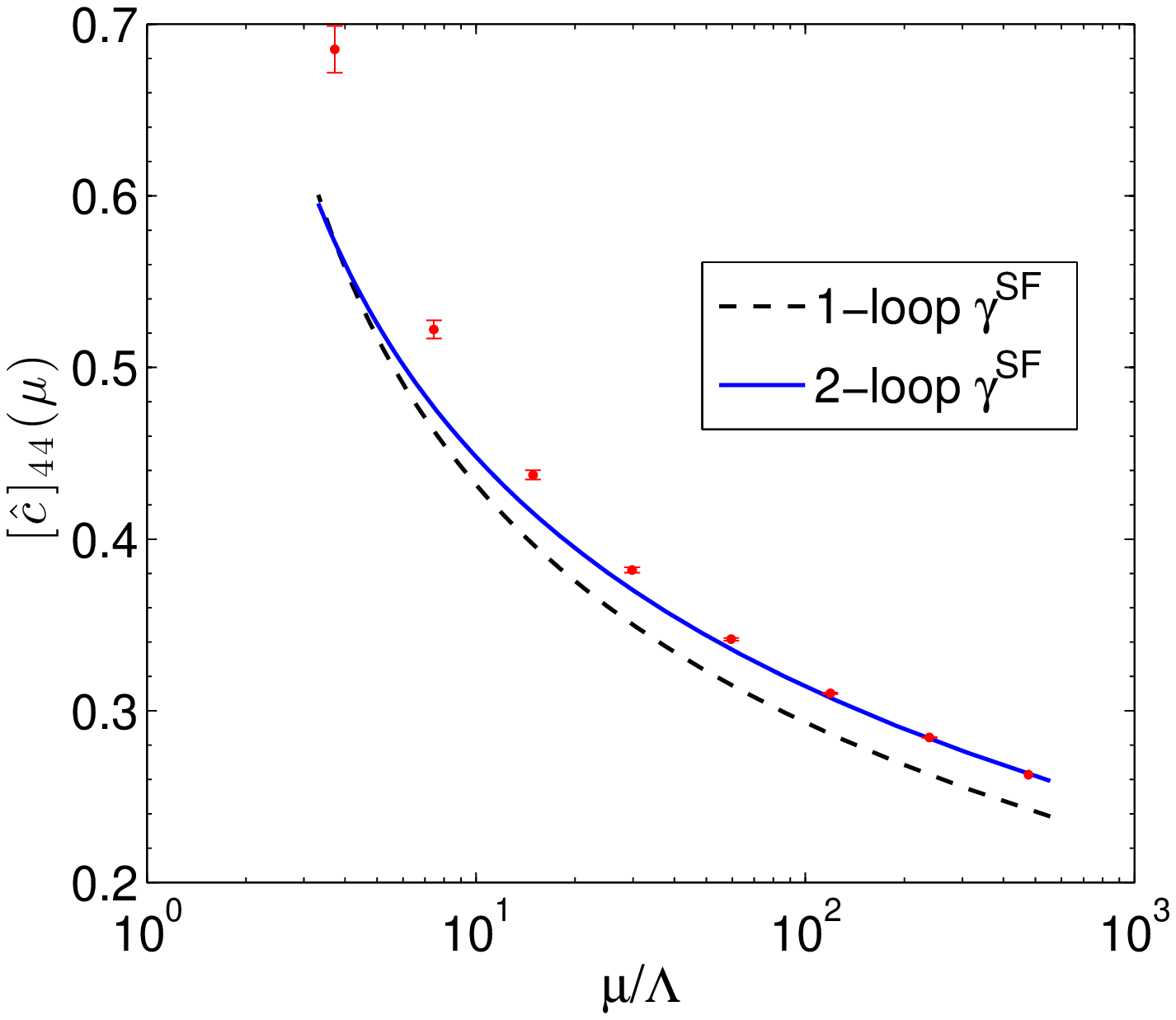}
      \hskip 0.0cm\includegraphics[width=5.1 true cm,angle=0]{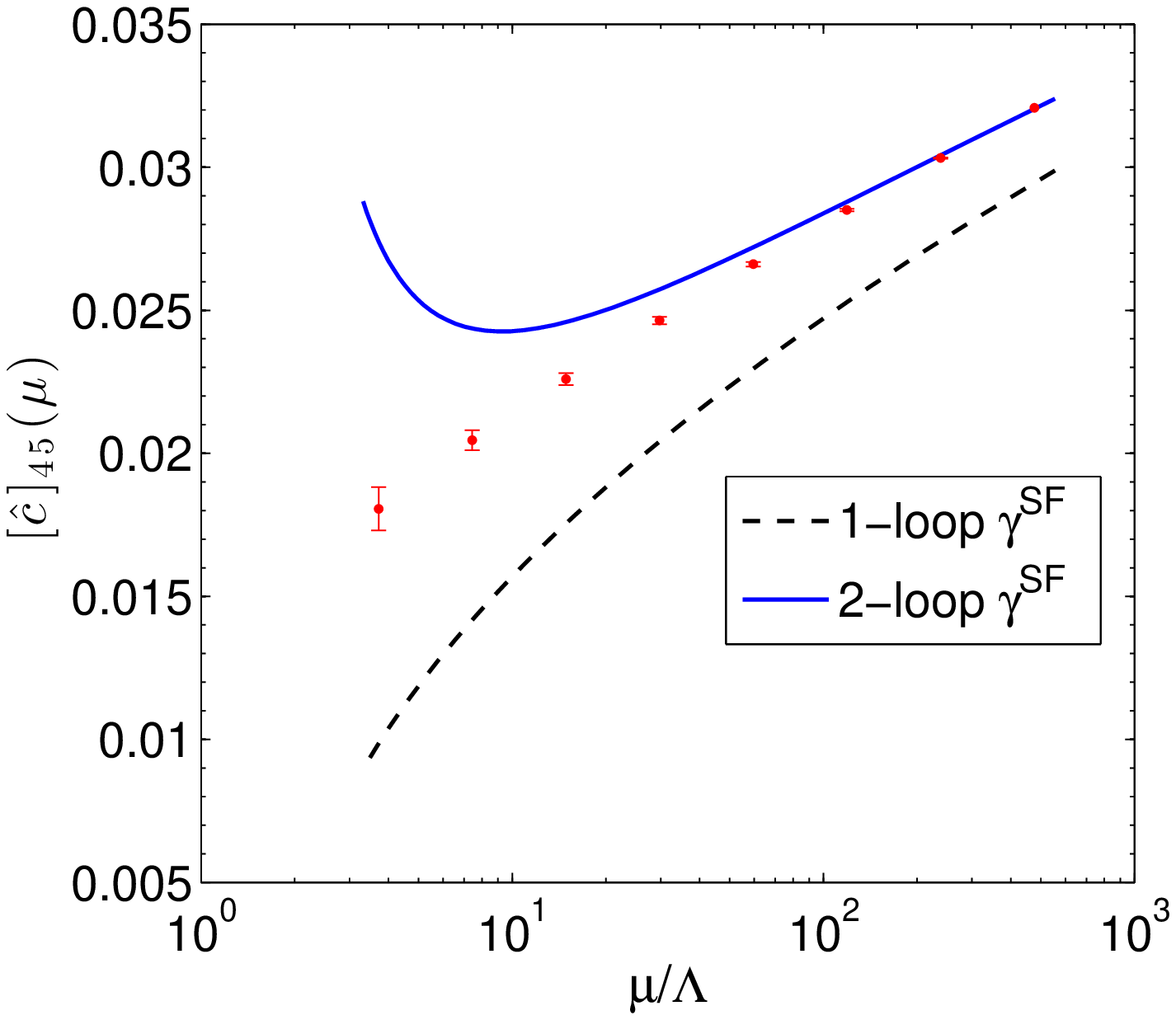}
      \hskip 0.0cm \includegraphics[width=5.1 true cm,angle=0]{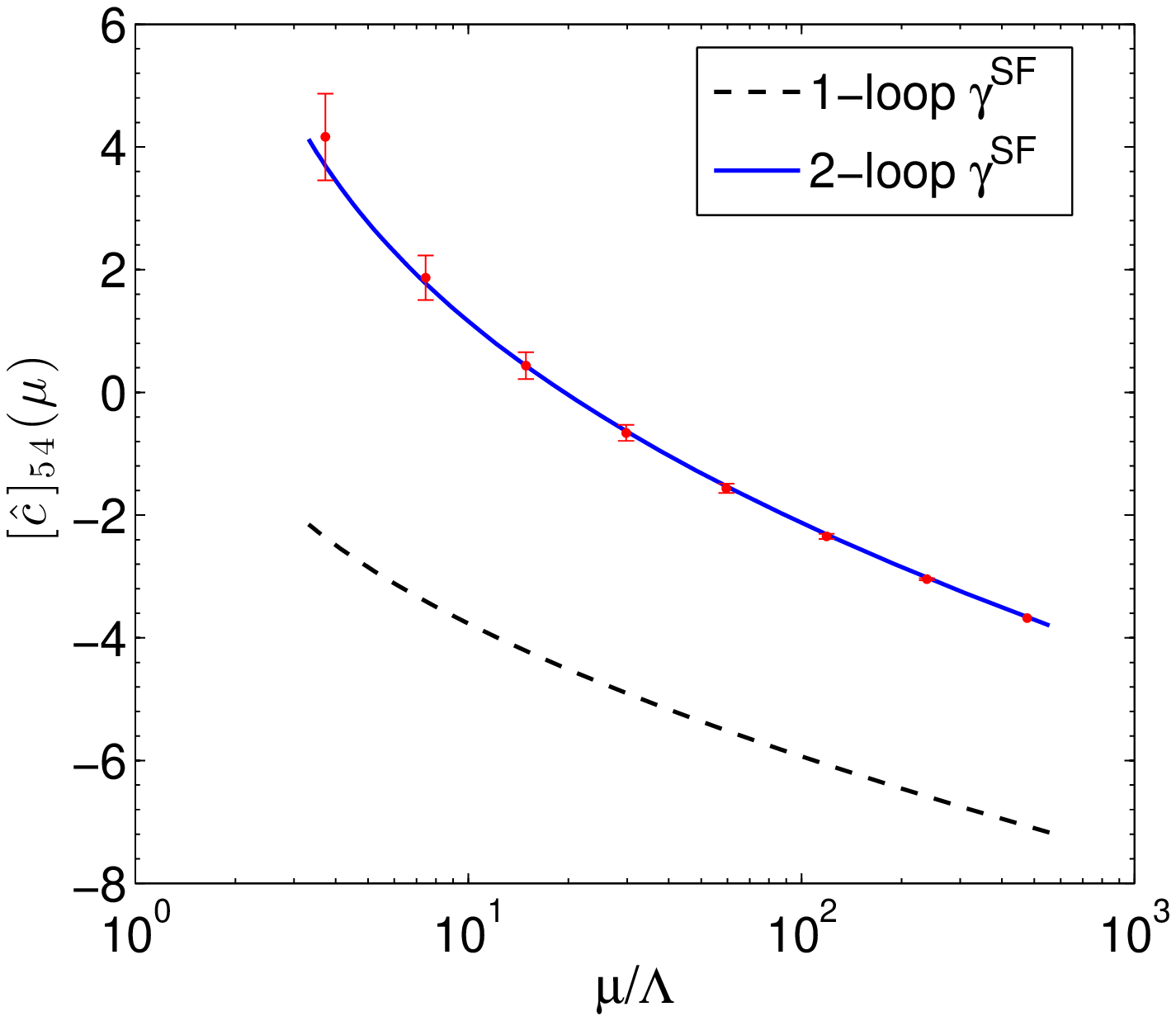}
       \vskip -0.3cm
    \caption{First row: elements 22, 23, 32 of $\tilde U(\mu)$. Second row: 
    elements 44, 45, 54 of $\tilde U(\mu)$.}
    \label{fig:RGrun}
  \end{center}
\vskip -0.6cm
\end{figure}

If operators mix, the RG-evolution is formally obtained by using
\beq
U(\mu_2,\mu_1)={\rm T}\exp\left\{\int_{\gbar(\mu_1)}^{\gbar(\mu_2)}\frac{\gamma(g)}{\beta(g)} {\rm d}g\right\}.
\eeq
We write the RG-evolution by separating the LO part and defining the function $W(\mu)$ which can be thought as containing contributions beyond LO:
\beq
U(\mu_2,\mu_1)\equiv[W(\mu_2)]^{-1}U(\mu_2,\mu_1)_{\rm LO}W(\mu_1)\ ,\qquad\qquad U(\mu_2,\mu_1)_{\rm LO}= \left[\frac{\gbar^2(\mu_2)}{\gbar^2(\mu_1)}\right]^{\frac{\gamma_0}{2\beta_0}}\ ,
\eeq
where $W(\mu)$ satisfies a new RG-equation and is regular in the UV: $\lim_{\mu\rightarrow\infty}W(\mu)=\mathbf{1}$ 

The RGI operators are easily defined using the above form:
\beq
\cQ^{\rm RGI}\equiv{\tilde U(\mu)} (\cQ(\mu))_{\rm R}=\left[\frac{\gbar^2(\mu)}{4\pi}\right]^{-\frac{\gamma_0}{2\beta_0}}W(\mu)(\cQ(\mu))_{\rm R}\ .
\label{eq:RGI1}
\eeq

This formula is still valid non-perturbatively. One can use it to perform the matching at $\mu_{\rm pt}$ with the NLO perturbative evolution:
\beq
\cQ^{\rm RGI}=\left[\frac{\gbar^2(\mu)}{4\pi}\right]^{-\frac{\gamma_0}{2\beta_0}}W(\mu_{\rm pt})U(\mu_{\rm pt},\mu_{\rm had})(\cQ(\mu_{\rm had}))_{\rm R}\ ,
\label{eq:RGI2}
\eeq
by expanding $W(\mu)$ in perturbation theory $W(\mu)\simeq\mathbf{1}+\gbar^2(\mu)J(\mu)+O(\gbar^4)$. $J$ depends on the ADM at the NLO $\gamma_1$ and satisfies:
\beq
\frac{\partial}{\partial\mu}J(\mu)=0\ ,\qquad J-\left[\frac{\gamma_0}{2\beta_0},J\right]=\frac{\beta_1}{2\beta_0^2}\gamma_0-\frac{1}{2\beta_0}\gamma_1\ .
\label{eq:J}
\eeq
Eq.~\ref{eq:J} has been solved to obtain $J$ in the SF scheme and compute the running $\tilde U(\mu)$ defined by Eq.~\ref{eq:RGI1},\ref{eq:RGI2}. As an example, results for some elements of $\tilde U(\mu)$ in the $\alpha=3/2$, $(s_1,s_2)=(1,5)$ scheme are presented in Fig.~\ref{fig:RGrun} against the LO and the NLO perturbative results. Again, in general several of these elements differ substantially at the lowest scales from the NLO PT results, independently of the scheme chosen.

The total RGI renormalization matrix is defined from $\cQ^{\rm RGI} \equiv \cZ^{\rm RGI}(g_0)\cQ(g_0)$ where 
\beq
\cZ^{\rm RGI}(g_0)=\tilde U(\mu_{\rm pt}) U(\mu_{\rm pt},\mu_{\rm had}) {\cZ}(g_0,a\mu_{\rm had})
\eeq
and $\cZ(g_0,a\mu_{\rm had})$ is the non-perturbative renormalization constant matrix computed at the hadronic scale.  $\cZ(g_0,L/a)$ has been computed at three values of $\beta\in\{5.20,5.29,5.40\}$ useful for large volume simulations, on three volumes for each $\beta$ ($L/a=\{4,6,8\}$). By interpolating to $L_{\rm max}$ for which $\gbar^2(L_{\rm max})=4.61$ one gets $\cZ(g_0,a\mu_{\rm had})$ for each $\beta$. 

 
\section{Conclusions}

Thanks to the use of SF schemes, we have performed a first exploratory study of the non-perturbative  RG-running of four-quark operators in the presence of mixing on a wide range of scales which varies over 2 orders of magnitudes.
Non-perturbative effects seem dangerously sizeable at scales of 2-3 GeV. Despite the dependence 
on the scheme, we are trying to understand whether this observation can be at the origin of the discrepancies found in the literature~\cite{Boyle:2012qb,Bertone:2012cu,Bae:2013tca} where the NLO perturbative RG-running in the RI-MOM or $\MSbar$ scheme have been used. The same strategy used here is immediately portable to $N_f=2+1$ dynamical simulations. Moreover, by using the $\chi$SF scheme~\cite{Sint:2010eh} one would gain automatic $O(a)$ improvement and only need 3-point functions instead of 4-point functions, with a consequent reduction of statistical fluctuations.

\end{document}